\documentclass{aa}
\usepackage{psfig}
\def\teff{\ifmmode T_{\rm eff} \else $T_{\mathrm{eff}}$\fi}

\def\ltsima{$\buildrel<\over\sim$}
\def\lsim{\lower.5ex\hbox{\ltsima}}

\newcommand{\hii}{H~{\sc ii}}
\newcommand{\ha}{\ifmmode {\rm H}\alpha \else H$\alpha$\fi}
\newcommand{\hb}{\ifmmode {\rm H}\beta \else H$\beta$\fi}
\newcommand{\lya}{\ifmmode {\rm Ly}\alpha \else Ly$\alpha$\fi}

\newcommand{\Heiiuv}{He~{\sc ii} $\lambda$1640}

\newcommand{\ltapprox}{\raisebox{-0.5ex}{$\,\stackrel{<}{\scriptstyle\sim}\,$}}
\newcommand{\gtapprox}{\raisebox{-0.5ex}{$\,\stackrel{>}{\scriptstyle\sim}\,$}}

\def\micron{$\mu$m}
\def\kms{km s$^{-1}$}

\def\ergscm{erg s$^{-1}$ cm$^{-2}$}
\def\msun{\ifmmode M_{\odot} \else M$_{\odot}$\fi}
\def\zsun{\ifmmode Z_{\odot} \else Z$_{\odot}$\fi}
\def\lsun{\ifmmode L_{\odot} \else L$_{\odot}$\fi}

\def\mup{\ifmmode M_{\rm up} \else M$_{\rm up}$\fi}
\def\mlow{\ifmmode M_{\rm low} \else M$_{\rm low}$\fi}

\def\aap{A\&A}
\def\aaps{A\&AS}

\def\aj{AJ}
\def\apj{ApJ}

\def\apjs{ApJS}
\def\mnras{MNRAS}

\newcommand{\oh}{\ifmmode 12 + \log({\rm O/H}) \else$12 + \log({\rm
O/H})$\fi}
\newcommand{\nii}{[N~{\sc ii}]}

\def\Oii{[O~{\sc ii}] $\lambda$3727}
\def\Oiii{[O~{\sc iii}] $\lambda\lambda$4959,5007}

\begin{document}
\title{ISAAC/VLT observations of a lensed galaxy at $z=10.0$
\thanks{Based on observations collected with the
ESO VLT-UT1 Antu Telescope (70.A-0355, 271.A-5013),
the Hubble Space Telescope ({\it HST})
and the CFHT.}}

\author{R.~Pell\'o\inst{1},
D.~Schaerer\inst{2,1},
J.~Richard\inst{1},
J.-F.~Le~Borgne\inst{1},
J.-P.~Kneib\inst{3,1}
}
\offprints{R.~Pell\'o, roser@ast.obs-mip.fr}
\institute{
Laboratoire d'Astrophysique (UMR 5572),
Observatoire Midi-Pyr\'en\'ees,
14 Avenue E. Belin, F-31400 Toulouse, France
\and
Observatoire de Gen\`eve,
51, Ch. des Maillettes, CH-1290 Sauverny, Switzerland
\and
Caltech Astronomy, MC105-24, Pasadena, CA 91125, USA
}
\date{Received 20 January 2004 / Accepted 13 february 2004}
\authorrunning{R.~Pell\'o et al.}{ }
\titlerunning{ISAAC/VLT observations of a lensed galaxy at $z=10.0$}{}

\abstract{We report the first likely spectroscopic confirmation 
of a $z\sim 10.0$ galaxy from our ongoing search for distant
galaxies with ISAAC/VLT.
Galaxy candidates at $z \gtapprox 7$ are selected from
ultra-deep $JHKs$ images in the core of gravitational lensing
clusters for which deep optical imaging is 
also available, including HST data. 
The object reported here, found behind Abell 1835, 
exhibits a faint emission line
detected in the $J$ band, leading to $z=10.0$ when identified as \lya,
in excellent agreement with the photometric redshift determination. 
Redshifts $z < 7$ are very unlikely for various reasons we 
discuss.
The object is located on the critical lines corresponding to z$=$9 to 11.
The magnification factor $\mu$ ranges from 25 to 100.
For this object we estimate
$SFR(\lya) \sim  (0.8-2.2)$ \msun yr$^{-1}$ and $SFR(UV) \sim (47-75)$
\msun\ yr$^{-1}$, both uncorrected for lensing. The steep UV slope
indicates a young object with negligible dust extinction. 
SED fits with young low-metallicity stellar population models
yield (adopting $\mu=25$)
a lensing corrected stellar mass of $M_\star \sim 8 \times 10^6$ \msun,
and luminosities of $2 \times 10^{10}$ \lsun, corresponding to a dark
matter halo of a mass of typically $M_{\rm tot} \ga 5 \times 10^8$ \msun.
In general our observations show that under excellent conditions 
and using strong gravitational lensing direct observations of 
galaxies close to the ``dark ages'' are feasible with ground-based 
8-10m class telescopes.

}

\maketitle
\section{Introduction}
 Spectacular progress during the last decade has permitted direct observations 
of galaxies and quasars out to redshifts $z \sim 6.6$
(Hu et al.\ 2002, Fan et al.\ 2003, Kodaira et al.\ 2003), 
or in other words over more than 90 \% of cosmic time. 
With the exception of the cosmic microwave background, 
direct exploration of more distant 
objects has so far been hampered by the need for observations beyond the optical 
domain, and technical difficulties related to near-IR observations of such faint objects. 
However, by combining a strong magnification of background galaxies by well known 
gravitational lensing clusters with 
present-day near-IR instruments on 8-10m class telescopes such an attempt appears 
feasible (Schaerer \& Pell\'o\ 2002; Pell\'o et al.\ 2003). 
We present here the first spectroscopic results obtained from our
prototype observational program with ISAAC/VLT on one of our best $z
\ga 7$ galaxy candidates. 

In Sect.~\ref{observations} we summarize the observational strategy
adopted in the present study, and we present the photometric and
spectroscopic characteristics of this object.
The redshift identification is discussed in Sect.\ \ref{s_z}.
The physical properties of this galaxy and implications of our finding
are briefly discussed in
Sect.~\ref{results}.

\section{Strategy, observations, and redshift determination}
\label{observations}

\begin{figure}[htb]
\centerline{\psfig{figure=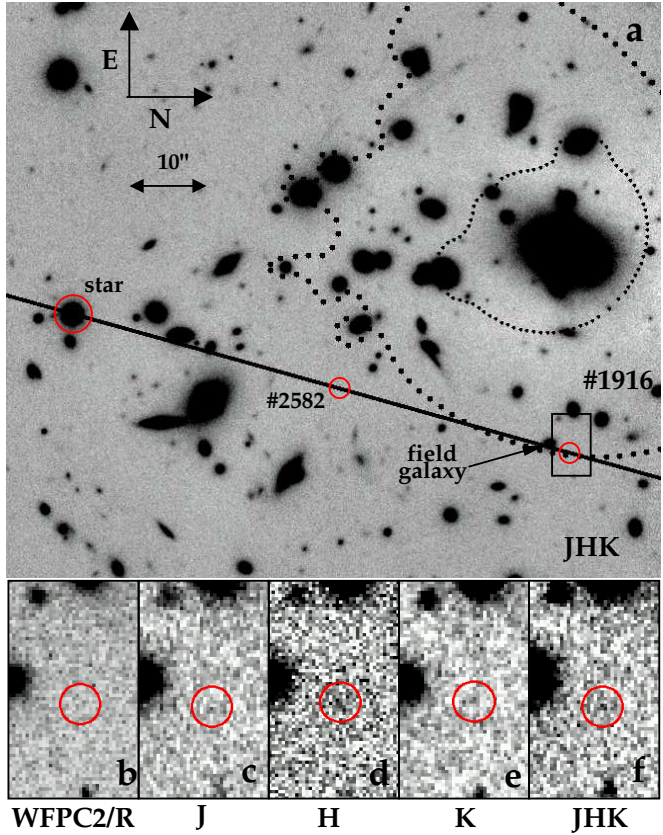,width=8.8cm}}
\caption{Composite $JHKs$ ISAAC image showing:
{\bf a)} The core of the lensing cluster A1835, with the position of
the long slit used during our spectroscopic survey with ISAAC/VLT,
together with the location of objects \#1916 and \#2582,
the reference
acquisition star (circles), and a nearby field galaxy seen on the 2D
spectra. Large and small dots show the position of the 
external and internal critical lines at $z=10$. 
{\bf b)} Thumbnail image in the 
HST-WFPC2 $R$ band, where the object is not detected. 
{\bf c, d, e)} $J$, $H$, and $Ks$ images respectively. 
{\bf f)} Composite $JHKs$  image. All thumbnail images are 
scaled to the original ISAAC pixel size, without smoothing, and are displayed 
with a linear scale. 
}
\label{fig1}
\end{figure}

\begin{figure}[htb]
\centerline{\psfig{figure=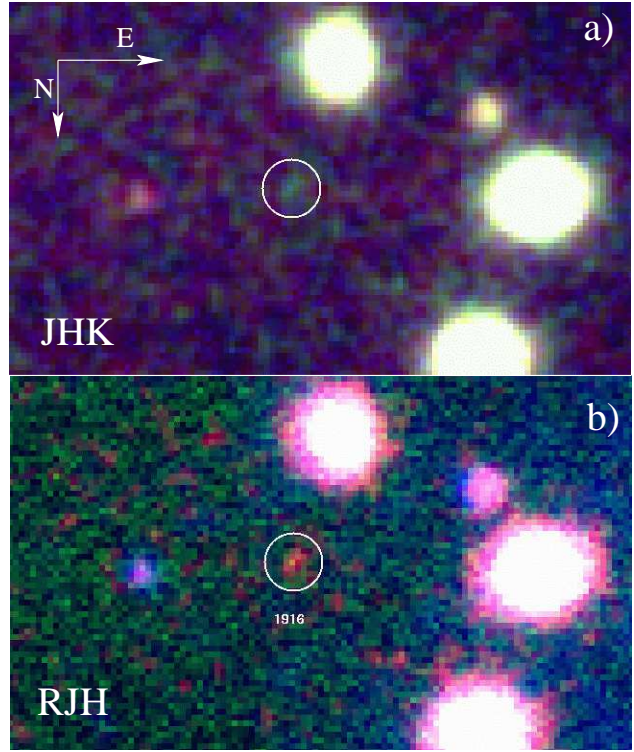,width=8.8cm}}
\caption{True color composite images displaying the color difference
between the lensed galaxy \#1916 and other objects in this field: 
{\bf a)} JHK and {\bf b)} RJH. This object is blue in JHK as compared
to cluster galaxies due to its intrinsically blue UV rest frame
spectrum in HK, whereas it is red in RJH due to the presence of a break
within the J band.
}
\label{fig1b}
\end{figure}

 To search for galaxy candidates at very high redshift ($z > 7$) we used the 
following strategy. We apply the traditional Lyman drop-out technique (Steidel et al.\ 1996)
to deep optical images.
At redshift $z \ga 6$ the main spectral discontinuity is due to the Gunn-Peterson 
(GP) trough, i.e. the nearly complete absorption of the flux shortward of \lya\ 
(1216 \AA), due to the large neutral H column density in the intergalactic medium (IGM). 
In practice we search for objects which are non-detected in optical ($VRI$) bands. In 
addition we require a detection in $J$, $H$, and $Ks$, a fairly red $J-H$ due the appearance of 
the GP trough in $J$, and a blue $H-Ks$ colour corresponding to an intrinsically blue UV 
restframe spectrum such as expected for UV bright starbursts. These criteria, devised 
from simulations based on synthesis models at different metallicities
(Schaerer \& Pell\'o 2002, Schaerer 2003, Pell\'o et al.\ 2003) allow us to 
select galaxies with redshifts between 7 and 10. Furthermore the detection in at least 
two bands longward of \lya\  and the combination with the above $H-Ks$ colour 
criterion allows us to avoid contamination by cool stars.

\subsection{Photometry}
   Ultradeep JHKs imaging of the central $2\times 2$ arcmin$^2$ of the lensing cluster  Abell 
1835 ($z=0.253$) was obtained at ESO/VLT with the near-IR imager and
spectrograph ISAAC in February 2003. 
The total exposure times and the -- excellent -- seeing conditions are given in
Table 1. Indeed, a large fraction of the total exposure time in H and
Ks was obtained under excellent seeing conditions, with 74 \% of the data
below $0.6''$ in H and 84 \% below $ 0.4''$  in Ks.
Photometric data were complemented by deep $VRI$ observations 
taken at the Canada-France Hawaii Telescope (see Table 1)
and $R$ band (F702W) WFPC2/HST images 
(Programme ID:8249, PI: Kneib, exposure time 7.5 ksec). 
Near-IR images were combined and stacked following a
procedure similar to the one used by Labbe et al.\ (2003),
which is well suited for point sources like the ones expected here.
Some modifications were adopted to improve the background subtraction within a
cluster of galaxies.
Photometry was performed on a 1.5 arcsec aperture
using the SExtractor package (Bertin \&
Arnouts 1996). Instead of adopting the SExtractor error bars, which
are based on sky noise only, we have computed more
realistic background noise statistics by randomly sorting 200 blank regions of
1.5 arcsec aperture on the neighbouring sky area $\sim$ 10 arcsec
around the object. The resulting error estimates (adopted in Table 1)
are larger by 0.1 mag ($J, H$) and 0.2 mag ($Ks$) than the SExtractor
values. 
The above procedure has also been used to derive the
random noise within $0.6''$ apertures used to constrain the limiting
magnitudes in the filters where the object is not detected.
We have also checked that all the optical and near-IR colors for
cluster galaxies and, in particular the $H-Ks$ versus $J-H$
color-color diagram, are fully compatible with the expected values for
$z=0.25$ galaxies.
A more detailed report of these observations and
analysis of the photometric data will be published elsewhere (Richard
et al.\, in preparation). 

\begin{table}
\caption{VRIJHKs photometry of \#1916. 
Near-IR magnitudes are measured within 
1.5 arcsec apertures on seeing matched images. 
Limiting magnitudes in the visible bands correspond to a 
detection limit of 1 $\sigma$ within the near-IR reference seeing
aperture, which is equivalent to a 3 $\sigma$ detection on 4 ISAAC pixels.
Table entries denote the filter name (col. 1), effective wavelength (2), total 
exposure time (3), the image quality/seeing (4), and the observed 
magnitudes in the Vega (5) and AB (6) systems.}
\begin{tabular}{lllllll}
\hline

Filter & $\lambda_{\rm eff}$ & $t_{\rm exp}$ & seeing & mag    & mag   \\
       &   [$\mu$m]          & [ksec]    & [$''$]    & [Vega] &  [AB] \\
\hline 
V       &0.54   &3.75   &0.76   &$>$ 28.5               &$>$ 28.6       \\
R       &0.66   &5.4    &0.69   &$>$ 28.5               &$>$ 28.7       \\
I       &0.81   &4.5    &0.78   &$>$ 27.6               &$>$ 28.1       \\
J       &1.26   &6.48   &0.65   &25.83 $\pm$ 1.00       &26.77 $\pm$ 1.00\\
H       &1.65   &13.86  &0.5    &23.59 $\pm$ 0.25       &25.00 $\pm$ 0.25\\
Ks      &2.16   &18.99  &0.38   &23.64 $\pm$ 0.36       &25.51 $\pm$ 0.36\\
\hline
\end{tabular}
\label{table1}
\end{table}

\begin{figure}[htb]
\centerline{\psfig{figure=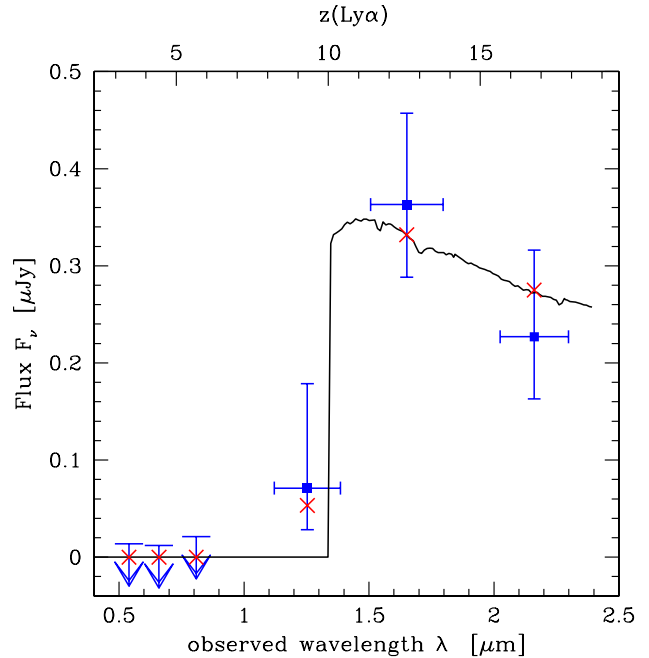,width=8.8cm}}
\caption{Optical to near-IR spectral energy distribution of the 
galaxy \#1916. Shown are the broadband photometric measurements and 
associated 1 $\sigma$ uncertainties for $JHKs$. Three $\sigma$ upper limits on 4 pixels are 
given for the non detections in $VRI$. The solid curve shows a good (98.7 \% 
probability) model fit with a synthetic spectrum of a young unreddened starburst 
with redshift $z=10.0$ of metallicity $Z=1/5 \zsun$, constant star formation, and an 
age of 3 Myr. The crosses indicate the corresponding broad band fluxes of the 
model. The upper axis shows the corresponding \lya\ redshift.}
\label{fig2}
\end{figure}

Applying the above
search criteria has provided six $z > 7$ galaxy candidates in the observed
field. Among them, the best candidate (called \#1916 hereafter) was
retained for subsequent follow-up spectroscopy. The coordinates of this
object are $\alpha(2000)=$14:01:00.06, $\delta(2000)=$+02:52:44.1. The location and
broad-brand spectral energy distribution (SED) of \#1916 are shown in
Figs.\ \ref{fig1} and \ref{fig2}; photometric data is given in Table 1. 
The radius of the circles in the thumbnail images in Fig.~\ref{fig1} is similar to 
the aperture used to compute magnitudes.
Limiting magnitudes adopted in Table 1 correspond to a
detection limit of 1 $\sigma$ within the near-IR reference seeing
aperture ($0.6''$), which is roughly equivalent to a detection limit
of 3 $\sigma$ within a region equivalent to 4 ISAAC pixels.
The source is not detected on the WPC2/HST F702W image with $R(F702W) >$ 27.3 mag 
at 2 $\sigma$ on 4 HST pixels.
Figure \ref{fig1b} shows the JHK and RJH true color
composite images displaying the color difference
between galaxy \#1916 and other objects in this field, in particular
cluster galaxies.

The broad band SED of \#1916 was used to constrain the source redshift
using our public photometric redshift code {\em Hyperz} (Bolzonella et al.\ 2000)
including numerous
empirical and theoretical spectral templates for both galaxies and AGNs,
and keeping internal extinction as a free parameter. The resulting
redshift probability distribution shows a clear maximum at redshift
$z_{\rm phot} \sim$ 9--11. This photometric redshift estimate is mainly corroborated
by the finding of a strong break of $\ga$ 3.1 to 3.7 AB mags between
$VRI$ and $H$, which has a high significance independently of the definition
used for the limiting magnitudes. The low $J$-band flux is
attributed to the presence of the GP trough located within this filter
at these redshifts.

\subsection{Spectroscopy}
To attempt a spectroscopic redshift determination we have observed \#1916 in the $J$ band 
with ISAAC. The observations were taken between 29 June and 3 July 2003, using a 1 
arcsec slit width in order to achieve an optimum positioning using bright sources. The 
results presented here were obtained again under excellent seeing conditions
ranging between 0.4 and 0.5 arcsec (about 3 pixels in the composite frames).
The source is not resolved in the 
spatial direction. Thus, we obtained seeing-limited spectra for this source with an 
equivalent spectral resolution $R \sim 5400$.

\begin{figure}[htb]
\centerline{\psfig{figure=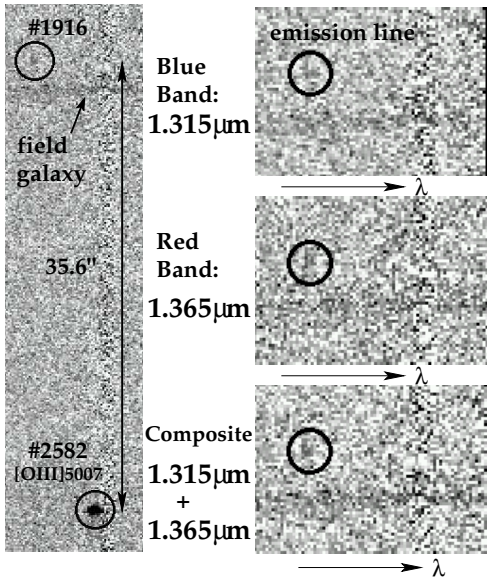,width=7cm}}
\caption{
2D spectra showing the detected emission line 
of \#1916, as well the nearby field galaxy and the 
[O~{\sc iii}] $\lambda$5007 line of the galaxy
\#2582 ($z=1.68$) used as a reference to stack the data sets. 
2D spectra are sky subtracted and show the spectral 
region around the emission-line at 1.33745 \micron, 
leading to $z=10.00175 \pm 0.00045$ when identified 
as \lya (1215.67 \AA). The line is seen on the 2 
independent overlapping bands at 1.315 and 1.365 \micron.
No smoothing was applied to these spectra. Both
the object \#2582 and the nearby field galaxy are 
identified in Fig.1.}
\label{fig4}
\end{figure}

\begin{figure}[htb]
\centerline{\psfig{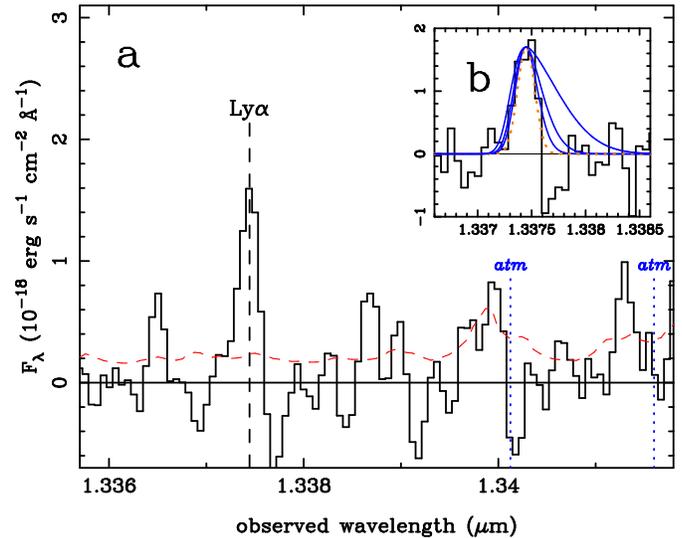}}
\caption{
{\bf a)} 1D spectrum of \#1916, extracted from the composite 2D
spectrum of the 1.315 and 1.365 \micron\ bands.
Dotted lines correspond to the 
position of the main sky OH emission lines. The continuum is not detected. The
dashed line displays the level corresponding to 1 $\sigma$ of the background noise. 
The displayed spectrum is smoothed to the instrumental resolution.
{\bf b)} Gaussian 
fit of the red half of the emission line, convolved with the instrumental profile 
(dotted line). Profiles corresponding to rest frame 60, 100 and 200 km s$^{-1}$ are 
also displayed for comparison (solid thick lines).}
\label{fig5}
\end{figure}

To search for 
possible faint emission lines, we have systematically explored the J band, where 
\lya\ should be located for objects within the $8 < z < 10$ redshift interval, with 
priorities set according to the photometric redshift probability distribution. For 
\#1916, the following effective exposure times were obtained in the different adjacent 
ISAAC bands: $\lambda_{\rm central} \sim$ 
1.193 \micron\ (3.6 ksec), 1.255 \micron\ (6.4 ksec), 
1.315 \micron\ (10.8 ksec) and 
1.365 \micron\ (18.9 ksec). Each band has  $\sim 600$ \AA\ width. 
Thus the observed spectral interval covers the range from 1.162 to 1.399 \micron;
in other words the ``surveyed'' interval corresponds to redshifts 
$z \sim$ 8.5--10.5 for \lya.
With a spectral resolution for the sky lines of 
$R=3100$, corresponding to the instrumental 1 arcsec slit width, the fraction of spectral band lost 
because of strong sky emission lines, mainly OH emission-lines, is of the order of 30\%.
Spectroscopic data were reduced using IRAF procedures and conforming to the ISAAC 
Data Reduction Guide 1.5\footnote{{\tt
    http://www.hq.eso.org/instruments/isaac/index.html}}, using the
same procedure described in Richard et al. (2003). 

The deep exposures resulted in the detection of one weak emission 
line at the 4-5 $\sigma$ level with an integrated flux of $(4.1 \pm 0.5)\times 10^{-18}$ 
erg cm$^{-2}$ s$^{-1}$ at a wavelength of 1.33745 \micron\ (see
Fig.\ \ref{fig4} and Fig.\ \ref{fig5}),
which appears on both the 1.315 and the 1.365 \micron\ bands. 
If identified as \lya\ the observed 
wavelength translates to a redshift of $z=10.00175 \pm 0.00045$
\footnote{Note that the given accuracy reflects only the line measurement assuming
a restframe $\lambda=1216$ \AA. Given the lack of knowledge of the intrinsic line profile
this represents a lower limit on the uncertainty on the true source redshift.}, 
in excellent agreement with 
the photometric redshift determination.

In the 1.315 \micron\ domain, we have also detected 3 emission lines on a secondary target
(called \#2582)
included in the same slit as \#1916 (Richard et al.\ 2003). 
This galaxy, a faint $R$-band dropout at $z=1.68$, 
shows \Oiii\ \AA\ and \hb. The weakest of them, \hb, has a flux 
of $(6.6 \pm 1) \times 10^{-18}$ \ergscm. 
This provides an independent verification that emission 
features as faint as $5 \times 10^{-18}$ \ergscm\ could have been
detected in our spectra all the 
way from 1.25 \micron\ to the end of the $J$ band (1.399 \micron), with some gaps due to sky 
features. At wavelengths below 1.25 \micron, the detection efficiency becomes smaller. For an 
unresolved line of  $5 \times 10^{-18}$ \ergscm\ a S/N ratio between 2
and 3.5 (from blue to red) 
would be expected at $\lambda < 1.25$ \micron. If present, an unresolved line of 
10$^{-17}$ \ergscm\ would have been detected, with a S/N of 3.6 to 5.5, all the way from 
1.162 to 1.399 \micron. 

We used the good S/N of the [O~{\sc iii}] $\lambda$5007 line of the galaxy
\#2582 at $z=1.68$, which is seen in both the 1.315 and 1.365 \micron\
overlapping bands, to compute the offsets required to
stack both data sets. The reference star in the same slit, together with a
standard star, were used to carefully model the telluric absorptions
in these spectra. The correction applied on the relevant wavelength
domain is smooth. Fig.~\ref{fig4} displays the 2D spectra obtained on
the two different overlapping and independent bands.
The combined 1D extracted spectrum around the emission-line is 
shown in Fig.~\ref{fig5}.

\section{On the redshift of \#1916}
\label{s_z}

How plausible is the observation of \lya\ from such distant an object?
Despite the presence of neutral HI in the IGM at very high z, all galaxies with $z \sim 6-6.6$
have been found through their \lya\ emission and several explanations exist for a 
partial transmission of \lya\ emission through the ambient ISM
and intervening IGM (Haiman 2002, Santos 2003) which are also applicable to higher redshifts.
In \#1916  the comparison of UV and \lya\ luminosities indicates
such a possible partial transmission (see Sect.\ 4).
The observed line is hardly resolved and barely 
asymmetric (see Fig.~\ref{fig5}). Best fits are obtained for rest
frame widths below 60 km s$^{-1}$ using a simple 
gaussian to fit the red half of the \lya\ emission convolved with the instrumental 
profile. Line widths beyond 200 km s$^{-1}$ (similar to those of the sample of $z \sim 5.7$
galaxies of Hu et al.\ 2003) are excluded with a high confidence level, 
but intermediate widths of the order of 100 km s$^{-1}$ cannot be excluded because 
of the low S/N ratio in the wings of the line. 
In any case, starburst galaxies show in general a wide range of \lya\ profiles, 
especially objects which are not selected by their line emission, like \#1916. 
In our case, a narrower \lya\ profile could also be related to the rather small
inferred dark matter halo mass (cf.\ Sect.\ 4), corresponding to circular 
velocities of the order of $v_c \sim 10$ \kms.
Small or negligible line asymmetries as observed here can be explained 
within simple models by a variety of combinations of parameters including 
the intrinsic emission line width, the size of the surrounding \hii\ region, 
and galactic winds (cf.\ Haiman 2002, Santos 2003).
We therefore conclude that
the detection of \lya\ emission from such a high redshift object and its
observed profile are not unexpected.

How secure is the identification of the single line as \lya, i.e.\
the redshift $z=10.0$ ?
Although other line identifications corresponding to lower redshifts cannot be ruled out 
in principle, they appear very unlikely for various reasons.
\begin{itemize}
\item
First, from the constraints of the broad-band SED all solutions with 
$z \la 7$ have a smaller probability (see Schaerer et al.\ 2004 for
more details).
In fact possible low-$z$ ($z \sim 2.5$) solutions correspond either
to young bursts with implausibly large extinction ($A_V \ga$ 2--3
mag) necessary to supress the flux blueward of the Balmer break, 
or to old stellar populations, for which, however, no emission lines
are expected. 
For low-$z$ solutions invoking a Balmer break the single 
strongest emission line often seen in the corresponding wavelength 
domain is [O~{\sc ii}] $\lambda\lambda$3726.0,3728.8. 
However, this identification is excluded as the doublet 
would easily be resolved in our observations. 
Indeed, if the line was \Oii\ observed at z=2.59, it would appear
as a double line with the 2 peaks separated by $\sim 16$ pixels, 
the two profiles overlapping even for moderate velocity dispersions 
($\sim 50 km/s$). However, this is not observed. 
Other obvious solutions such as \ha\ or \nii\ at z=1.03 are
difficult to reconcile with the photometric SED
and the absence of other lines in this interval.

A rather ``exotic'' solution at $z \sim 2.5$, albeit of lower probability, 
can be found with an empirical template of a metal-poor blue compact galaxy
with strong emission lines such as SBS 0335-052. 
In this case the observed strong $H$ band flux is dominated by \Oiii, \hb, and
H$\gamma$ emission lines, whereas the spectral range covered by $Ks$ is 
devoid a strong lines. 
However, to decrease sufficiently the rest-frame UV flux between $J$ and $VRI$
an exceptionally large extinction of $A_V \sim 3.6$ is required.
In such a case the observed emission line should correspond to lines between
\Oii\ and $\sim$ 4000 \AA, including e.g.\ relatively strong lines
of [Ne~{\sc iii}] $\lambda$3868 or [Ne~{\sc iii}]+H7 $\lambda$3968
(cf.\ Izotov \& Thuan 1998). Whereas \Oii\ can be excluded (cf.\ above),
other identifications could be possible. However, more than one emission 
line could then possibly have been detected in the observed wavelength 
range (cf.\ above). Although quite unusual, such an explanation has
the advantage of being easily testable, as several very strong emission
lines would be expected in the $H$ band.
 
For the reasons just discussed redshifts $z \la 7$ seem very unplausible.
For $z \sim$ 7--10 other possible identifications of lines sometimes observed in 
ehigh excitation objects and Lyman break galaxies include lines at $\lambda \la$
1900 \AA\ such as C~{\sc iii}] $\lambda$1909 \AA, \Heiiuv\ \AA, C~{\sc iv} $\lambda$1550,
and N~{\sc v} $\lambda$1240 \AA. 
However, also in this case, more than one emission line could possibly 
have been detected in the observed wavelength range (cf.\ above).
 
\item The second argument in favour of our line identification is the 
consistency between the well defined photometric redshift and the 
spectroscopic measurement.

\item Finally, the location of our object on top of the critical 
lines at $z \sim 9-11$ (cf.\ Sect.\ 4) represents an additional 
element supporting our redshift identification. It is worth noting 
that this fact was not considered as a selection criterion.
\end{itemize}

For these reasons we consider for now the identification of the observed 
line as \lya\ as the most likely one.
Obviously future detections of other emission lines and possible 
imaging and morphological constraints on our lensing prediction would 
represent an important verification of our redshift determination.
Also additional very deep imaging (e.g.\ at 1 \micron) could help 
improving further the photometric redshift of \#1916 and of other 
candidates of our sample.

\section{Properties of the $z=10.0$ galaxy and implications}
\label{results}
  Assuming ``concordance'' cosmological parameters ($\Omega_m=0.3$, $\Omega_\Lambda=0.7$, and 
$H_0 = 70$ km s$^{-1}$ Mpc$^{-1}$) the age of the Universe at redshift 10.0 is just $\sim$  
460 Myr after the Big Bang. 
In other words the galaxy we have detected lies at a distance corresponding to 
only  $\sim$  4 \% of the current age of the Universe. 

     According to the lensing model for Abell 1835 (Smith et al.\
2002), within modelling errors the object \#1916 is located on top of
the critical lines at $z \sim 9-11$.  
Therefore the magnification factor $\mu$ is very large, 
ranging with a large uncertainty from 25 to 100, i.e.\ 3.5 to 5 mag! As in principle the 
likelihood to obtain the largest amplification is small we will adopt the minimum value 
of $\mu=25$. 
Once corrected for lensing, the 
intrinsic AB magnitude of \#1916 is $\ga$ 28.5 and 29 in the $H$ and $Ks$ bands respectively 
(cf.\ Table 1). Clearly, without strong gravitational lensing 
and excellent seeing conditions near-IR photometry and 
spectroscopy of such a faint source would be impossible with current 8-10m class 
telescopes. 

Because of its location close to the critical line the object 
must be multiply imaged 
by the cluster. However, the identification of the counter images is impossible with the 
present data. Deeper observations with the Hubble Space Telescope in the near-IR 
should allow to confirm our current interpretation. 
Redshift $\sim$ 7--10 sources are expected to have virial radius of the
order of $\ltapprox 1$ kpc, corresponding to $\ltapprox 0.1''$,
and thus they are unresolved on our present images. HST images with
diffraction-limited resolution should thus allow to confirm the high-z
nature of this source (e.g. by the detection of strong tangential magnification or 
multiple images), to constrain the lensing configuration through
morphological considerations, and to determine the physical scales
involved in star formation activities at such redshifts. 

     Estimates of the star formation rate (SFR) can be obtained from the \lya\ 
luminosity, the UV restframe flux, or SED fits. For a ``standard'' Salpeter IMF from 
1--100 \msun\ and constant star formation at equilibrium we find a lower limit of 
$SFR(\lya) \sim  (0.8-2.2)$ \msun yr$^{-1}$ and $SFR(UV) \sim (47-75)$ \msun\ yr$^{-1}$, 
both uncorrected for lensing, adopting published conversion factors (Schaerer 2003,
Kennicutt 1998). The lower SFR 
derived from \lya\ is thought to reflect in large part the effects of photon 
``destruction'' on the blue side of the \lya\ emission line by scattering in the IGM 
and losses due to other possible effects (dust, ISM geometry). SED fits with constant 
star formation rate (e.g.\ shown in Fig.\ \ref{fig2} for $Z=1/5 \zsun$ and 3 Myr age) yield typically 
$SFR \sim 20-120$ \msun\ yr$^{-1}$  (0.8 to 4.8 \msun\ yr$^{-1}$ after lensing correction) 
depending on the age after the onset of star formation. 
Lower SFR values would be obtained for ``non standard'' IMFs favouring massive stars.

     In general the shape of the overall UV SED does not allow one to distinguish age, 
metallicity, and extinction as these parameters alter the UV spectrum in a degenerate 
way. In the present case the situation is somewhat different, as the UV restframe 
spectrum is very blue when compared to other known starbursts (Heckman et al.\ 1998) and to model 
predictions (Leitherer et al.\ 1999, Schaerer 2003) --
the UV slope $\beta$ measured from the $H$ and $Ks$-band flux is 
$\beta \sim  -3.8 \pm 2.1$. 
Comparisons with observed UV spectra of galaxies and model predictions also show 
that the observed UV is unlikely to be significantly influenced by line emission (if 
present). From this we can quite safely conclude that the extinction must be small in this 
$z=10.0$ galaxy. 

     SED modeling also allows us to estimate the total stellar mass and luminosity 
involved in this star forming event. Scaling young burst models with metallicities 
$Z \ga 1/50 \zsun$ to the observed SED we obtain $M_\star \sim  2 \times 10^8$ \msun\
(or lensing corrected values of $M_\star \sim 8 \times 10^6$ \msun) and luminosities of 
$L  \sim  4 \times 10^{11}$ \lsun\ ($2 \times 10^{10}$ \lsun\ lensing corrected) 
for the above Salpeter IMF. Larger masses are obtained for older bursts. The 
total current stellar mass estimated for constant star formation over $\sim$ 3 Myr is in good 
agreement with the above estimate. Note that, as star formation is likely to continue for 
longer time, the total mass of stars to be formed during this starburst event is likely 
larger than the ``current'' stellar mass.

     The above estimates show that \#1916 is comparable to or somewhat heavier than 
the most massive old Globular Clusters or the most massive super star cluster observed 
in nearby starbursts (Mandushev et al.\ 1991, Ho et al.\ 1996, Mengel et al.\ 2002). 
Assuming typical values for the baryonic/dark matter content 
and for the star formation efficiency ($\Omega_b/\Omega_m \times f_{\rm SF} \sim 0.015$) 
this translates to a total mass of the dark matter halo of $M_{\rm tot} \ga 5 \times 10^8$  
\msun. Such ``massive'' halos correspond to the collapse of $\ga$ 2 $\sigma$ fluctuations 
at redshift $z \sim 10$, where, if metals are absent or very 
deficient, star formation is expected to occur thanks to cooling by atomic hydrogen
(Tegmark et al.\ 1997, Barkana \& Loeb 2001). 
``Massive'' starbursts like the one found here should contribute to the cosmic 
reionization; they are possibly even the dominant contributors (Ciardi et al.\ 2003).

In short, the observed and derived properties of \#1916 agree well with 
expectations of a young protogalaxy experiencing a burst of star formation at $z=10$. 
If we are witnessing the first star formation event, therefore potentially the formation of a 
massive primordial (so-called Population III) star cluster, cannot be established or rejected 
from the present data. Additional spectroscopic observations searching for 
other possible emission line signatures (e.g. \Heiiuv, C~{\sc iv} $\lambda$1548+51 \AA\
or other metal lines) are necessary to answer this question. 
In fact a measurement of such lines, although presumably intrinsically fainter than 
\lya, could well be feasible since the observed (``transmitted'') \lya\ flux
represents only a relatively small fraction of the flux emitted from the source
(cf.\ above). The expected flux in other lines could therefore become comparable 
to the observed emission line flux.

Finally, our observations show that under excellent conditions and using strong
gravitational lensing direct observations of galaxies close to the ``dark ages''
are possible with ground-based 8-10m class telescopes. We are looking forward 
to the exploration of this yet unknown territory from the ground and with the
forth-coming James Webb Space Telescope.

\acknowledgements

We are grateful to R. Behrend, B. Ciardi, T. Contini, A. Ferrara, 
Y. Izotov, M. Lemoine-Busserolle, D. Pfenniger, D. Valls-Gabaud,
S. White and various other colleagues for useful comments and 
discussions. C. Charbonnel, G. Schaerer also contributed indirectly 
to the finalisation of this work.
We thank the ESO Director General for a generous allocation of
Director's Discretionary Time for ISAAC spectroscopy (DDT 271.A-5013).
Also based on observations collected at the European Southern 
Observatory, Chile (70.A-0355), the NASA/ESA Hubble Space Telescope 
operated by the Association of Universities for Research in Astronomy, Inc., 
and the Canada-France-Hawaii Telescope operated by the 
National Research Council of Canada, the French Centre National de la Recherche 
Scientifique (CNRS) and the University of Hawaii.
Part of this work was supported by the CNRS
and the Swiss National Foundation. JPK acknowledges support from
Caltech and the CNRS.


\end{document}